\documentclass[twocolumn,superscriptaddress,prl]{revtex4}
\usepackage{epsfig}
\usepackage{amssymb,amsfonts,amsmath}
\usepackage[pdftex]{pstricks}

\include{epsf}

\newcommand{\<}{\langle}
\renewcommand{\>}{\rangle}

\newcommand{\beq}{\begin{equation}}
\newcommand{\eeq}{\end{equation}}
\newcommand{\bea}{\begin{eqnarray}}
\newcommand{\eea}{\end{eqnarray}}

\usepackage{amsmath}
\usepackage{amssymb}
\usepackage{amsthm}
\usepackage{amsfonts}

\begin{document}
\title{Random versus maximum entropy models of neural population activity}
\author{Ulisse Ferrari}
\thanks{Equal contribution}
\affiliation{Institut de la Vision, INSERM and UMPC, 75012 Paris, France}
\author{Tomoyuki Obuchi}
\thanks{Equal contribution}
\affiliation{Department of Mathematical and Computing Science,
Tokyo Institute of Technology, Yokohama 226-8502, Japan}
\author{Thierry Mora}
\thanks{Corresponding author: tmora@lps.ens.fr}
\affiliation{Laboratoire de physique statistique, \'Ecole normale sup\'erieure, CNRS and UPMC, 75005 Paris, France}

\date{\today}
\linespread{1}

\begin{abstract}
The principle of maximum entropy provides a useful method for inferring statistical mechanics models from observations in correlated systems, and is widely used in a variety of fields where accurate data are available.
While the assumptions underlying maximum entropy are intuitive and appealing, 
its adequacy for describing complex empirical data has been little
studied in comparison to alternative approaches. Here data from the
collective spiking activity of retinal neurons is reanalysed. The
accuracy of the maximum entropy distribution constrained by mean
firing rates and pairwise correlations is compared to a random ensemble of
distributions constrained by the same observables. In general, maximum
entropy approximates the true distribution better than the typical or
mean distribution from that ensemble. This advantage improves with
population size,
with groups as small as 8 being almost always better described by
maximum entropy.
Failure of maximum entropy to outperform random models is found to 
be associated with strong correlations in the population.
\end{abstract}

\maketitle

The principle of maximum entropy was introduced in 1957 by Jaynes
\cite{Jaynes1957a,Jaynes1957} to formulate the foundations of
statistical mechanics as an inference problem. Its interest has been
recently rekindled by its application to a variety of data-rich
fields, starting with the correlated activity of populations of
retinal neurons \cite{Schneidman2006,Shlens2006}. The method has since
been used to study correlations in other neural data,
such as cortical networks \cite{Tang2008,Tavoni2015} and functional
magnetic resonance imaging \cite{Watanabe2013a}, as well as in other
biological and non-biological contexts, including multiple sequence
alignments of proteins \cite{Weigt2009,Mora2010,Figliuzzi2016} and
nucleic acids \cite{Santolini2014,DeLeonardis2015}, the collective
motion of bird flocks \cite{Bialek2012}, the spelling rules of words
\cite{Stephens2010}, and the statistics of decisions by the United
States supreme court \cite{Lee2015}. In many cases, the close link
between maximum entropy and statistical mechanics has led to new
insights into the thermodynamics of the system in terms of phase
transitions \cite{Mora2011a,Bialek2014,Tkacik2015}, or multi-valley
energy landscape \cite{Watanabe2014,Tkacik2014a}.
In other cases, the method has allowed for predictions of crucial practical relevance, such as residue contacts in proteins \cite{Morcos2011}, or deleterious mutations in HIV \cite{Ferguson2013}.

Although the motivations of maximum entropy
seem intuitive and can be formalized rigorously \cite{Shore1980},  the
perceived arbitrariness of its assumptions has led to question its
validity \cite{Aurell2016,VanNimwegen2016}. The starting point is to consider models that match empiral observations
on a few key statistics of the data. Maximum entropy's crucial---and arguably debatable---assumption is to pick, out of the many models that satisfy that constraint, the one with the largest Gibbs entropy. This choice seems natural, since it ensures that the model is as random as possible. However, it is not clear why it should describe the data better than other models satisfying the same constraints.
To address this question directly on empirical data, we reanalyse the original
neural data  from
\cite{Schneidman2006}, which contributed to the recent surge of interest in maximum entropy. We compare the accuracy of maximum entropy distributions to ensembles of distributions that satisfy the same constraints, using the approach developped in \cite{Obuchi2015a,Obuchi2015}.

\begin{figure}
\begin{center}
\noindent\includegraphics[width=\linewidth]{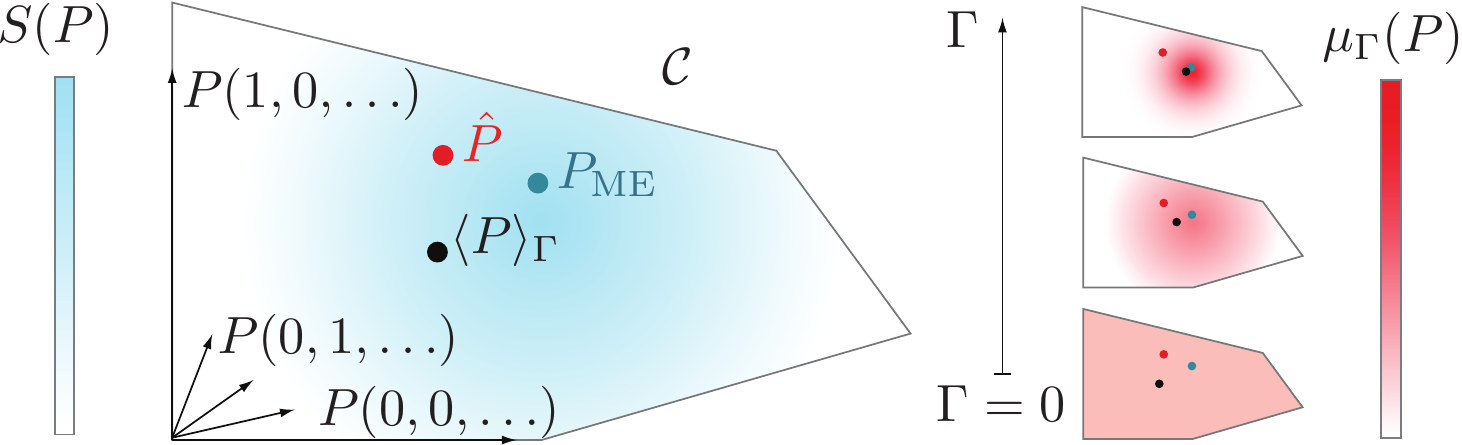}
\caption{{\bf Random models.} The space $\mathcal{C}$ of models $P$ is
  a simplex of $2^N-M-1$ dimensions, defined by the intersection of
  the hyperplanes satisfying the constraints that the mean observables
  under the model, $\sum_\sigma P(\sigma) \mathcal{O}_a(\sigma)$,
  $a=1,\ldots,M$, equal the empirical means,
  $\overline{\mathcal{O}}_a$, and by a normalization and positivity
  constraint. The true distribution to be approximated, $\hat P$ (red dot), is not accessible in general.
An entropy-dependent measure $\mu_\Gamma$  (Eq.~\ref{mu}) is defined on $\mathcal{C}$ (red map). At $\Gamma=0$ (random ensemble), the measure is uniform over that space.
As $\Gamma$ is increased, the measure concentrates onto the maximum entropy distribution $P_{\rm ME}$ (blue dot), and so does its mean $P_\Gamma=\<P\>_\Gamma$ (black dot).
\label{fig1}
}
\end{center}
\end{figure}

The collective state of a population of $N$ variables is described by 
$\sigma=(\sigma_1,\ldots,\sigma_N)$. 
In general, $\sigma_i$ may denote any degree of freedom, such as the identity of an amino-acid in a protein, the orientation of a bird in a flock, etc.
To fix idea, in this paper
$\sigma_i$ will be a binary variable describing the spiking activity of neuron $i$: $\sigma_i=1$ if neuron $i$ spikes within a given time window, and $0$ otherwise. The joint distribution of the collective activity $\sigma$, denoted by $P(\sigma)$, lives in a $2^N-1$ dimensional space, represented schematically in Fig.~\ref{fig1}. Because that space is huge for even moderately large populations, it is often impossible to sample the true distribution, $\hat P$, reliably from the data. Simplifying assumptions are needed.

To restrict the search of models, one can focus on distributions that
agree with the data on the average value of a few observables. Calling
these observables $\mathcal{O}_a(\sigma)$, $a=1,\ldots,M$, the
condition reads $P\cdot \mathcal{O}_a\equiv \sum_\sigma
P(\sigma)\mathcal{O}_a(\sigma)=\overline{\mathcal{O}}_a$, where
$\overline{\mathcal{O}}_a$ is the empirical mean. The observables must
be chosen carefully depending on the problem at hand, and may include
local or global order parameters, marginal probabilities, correlation
functions, etc. Let us denote by $\mathcal{C}$ the subspace of models
$P$ that satisfy those constraints, as well as the conditions
$P(\sigma)\geq 0$ and $\sum_\sigma P(\sigma)=1$. $\mathcal{C}$
is convex because of the linear nature of the contraints.

A probability law on $\mathcal{C}$ may be defined which weighs models
$P\in\mathcal{C}$ according to their Gibbs entropy,
$
S(P)=-\sum_{\sigma} P(\sigma)\log P(\sigma)$,
through the following measure \cite{Obuchi2015a}:
\beq\label{mu}
\mu_{\Gamma}(P)=\frac{e^{\Gamma S(P)}}{\mathcal{Z}},\qquad
\mathcal{Z}=\int_{P\geq 0} \mathcal{D}P\, e^{\Gamma S(P)},
\eeq
with
\beq
\mathcal{D}P= \delta\left(\textstyle{\sum_\sigma} P(\sigma)-1\right)\,
\prod_{a=1}^M \delta\left(P\cdot \mathcal{O}_a - \overline{\mathcal{O}}_a\right)\,\prod_{\sigma} dP(\sigma),
\eeq
where $\delta(\cdot)$ is Dirac's delta function.
The parameter $\Gamma$ is conjugate to the
entropy, and sets its average value: $\<S(P)\>_{\Gamma}=\partial \ln
\mathcal{Z}/\partial \Gamma$, where we use the brackets
$\<\cdot\>_{\Gamma}$ for averages over the measure $\mu_\Gamma$. $\Gamma$
plays the same role with respect to the entropy as the inverse
temperature with respect to the energy in standard statistical
mechanics. 
When $\Gamma=0$, all distributions satisfying the constraints have the
same probability. We call this the unbiased ensemble.
As $\Gamma\to\infty$, the measure becomes increasingly peaked onto a single
distribution, $P_{\rm ME}$, of maximum entropy (or, in the previous analogy,
the ground state reached at zero temperature). This distribution defines
 the classical
maximum entropy model, and takes the form \cite{Presse2013a}:
\beq\label{maxent}
P_{\rm ME}(\sigma)=\frac{1}{Z}\exp\left[\sum_{a=1}^M \lambda_a \mathcal{O}_a(\sigma)\right],
\eeq
where $\lambda_a$ are Lagrange multipliers enforcing the constraints on the mean observables, and $Z$ is a normalization constant.
We define the average distribution as
$P_{\Gamma}(\sigma)\equiv \<P(\sigma)\>_{\Gamma}$, which belongs to
$\mathcal{C}$ by convexity. 
$P_{\Gamma}$ only coincides with $P_{\rm ME}$ in the limit ${\Gamma\to\infty}$. At the other extreme, $P_0$ is the unbiased, center-of mass distribution that satisfies all the constraints.

\begin{figure}
\begin{center}
\noindent\includegraphics[width=.49\linewidth]{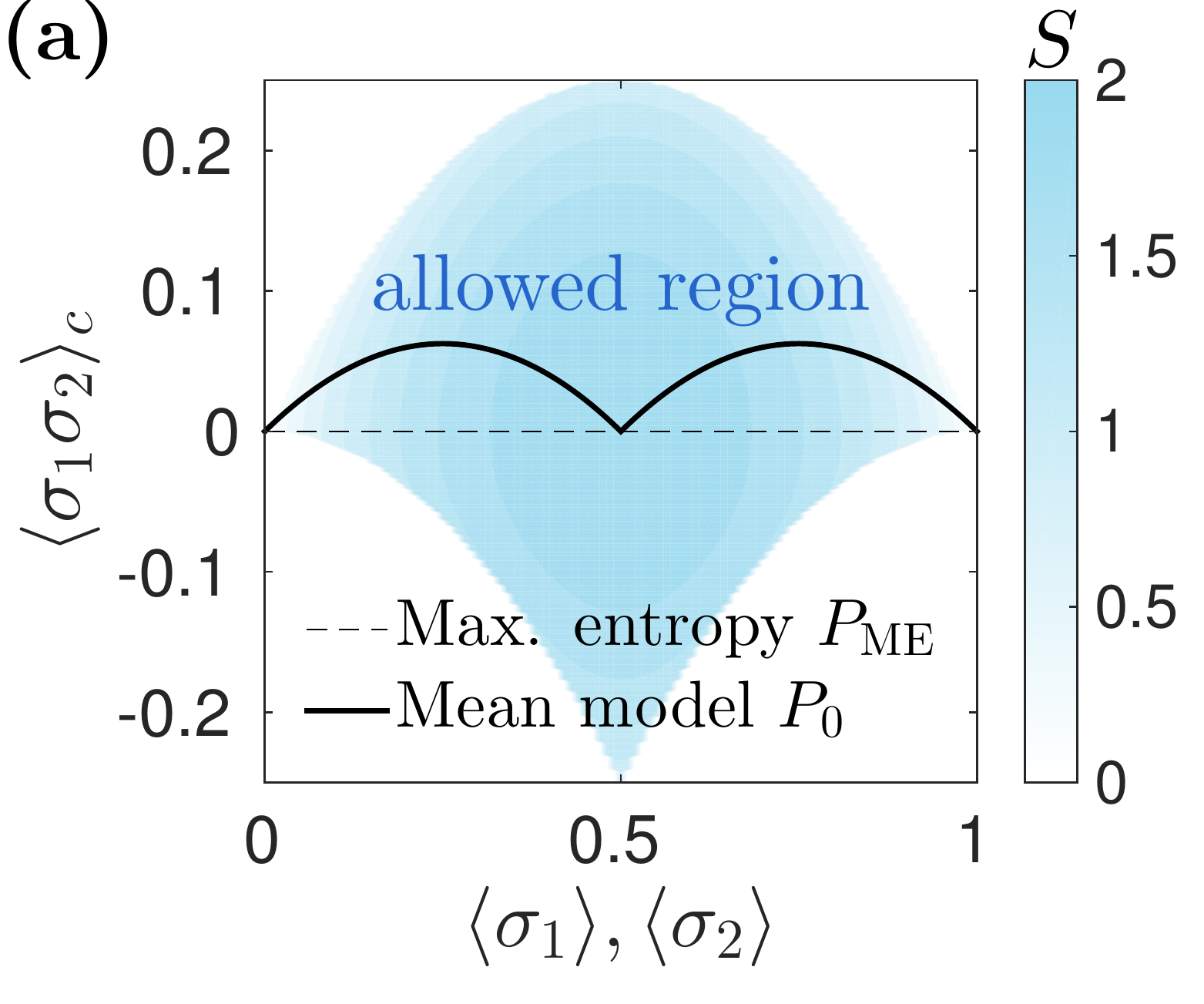}
\noindent\includegraphics[width=.49\linewidth]{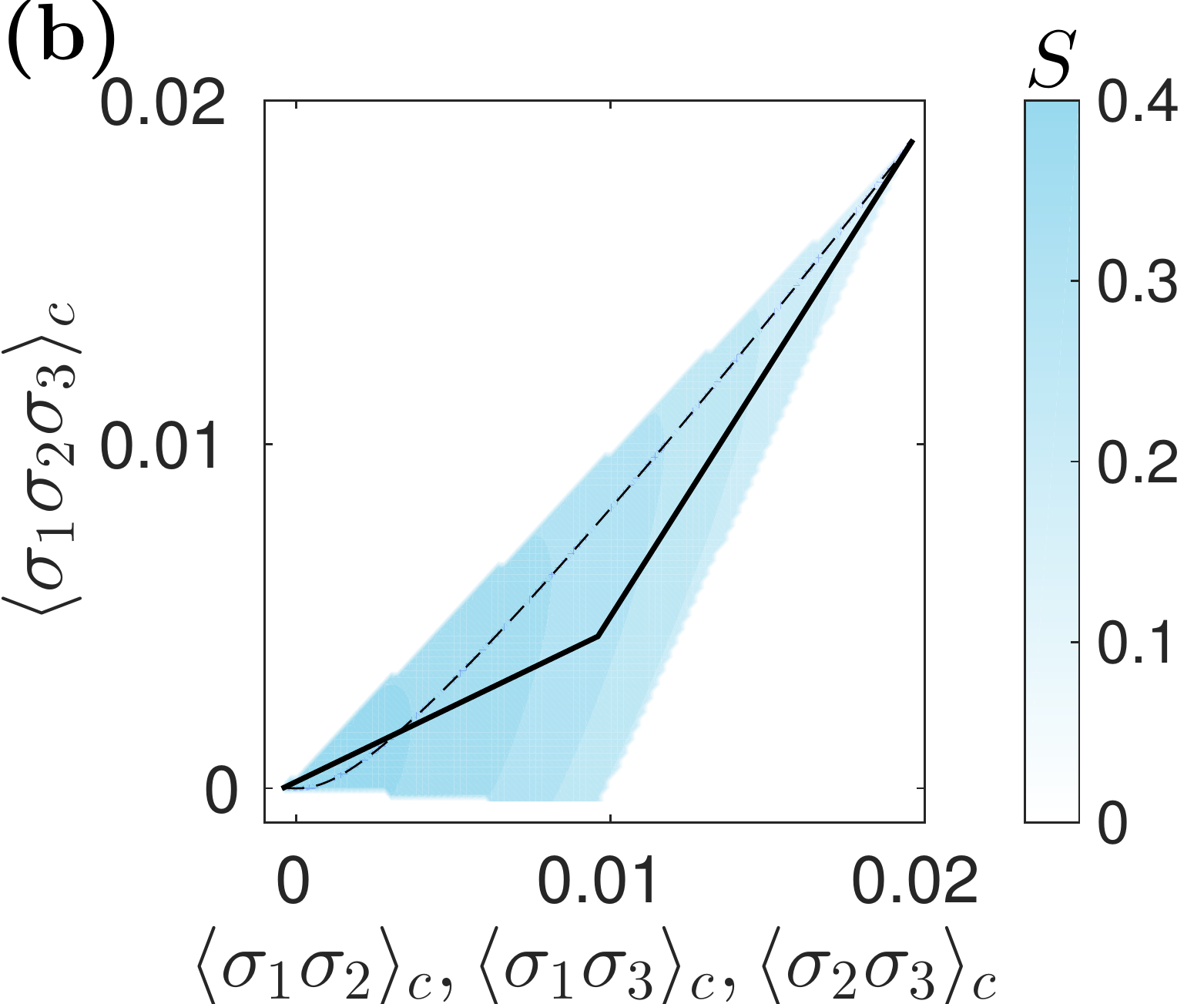}
\caption{{\bf Small networks.} Illustration of the random ensemble on
   2 and 3 neurons. (a) Pairwise correlation $\<\sigma_1 \sigma_2\>_c=\<\sigma_1 \sigma_2\>-\<\sigma_1\>\<\sigma_2\>$ predicted by maximum
   entropy and random models constrained by the mean spiking rates of
   two neurons, $\<\sigma_1\>=\<\sigma_2\>$, as a function of that
   rate. The mean unbiased model $P_0$ is the center of mass between
   the lower and upper allowed limits of the correlation, which
   delimit the shaded area. (b) Triplet connected correlation, $\<\sigma_1 \sigma_2 \sigma_3\>_c=\<\sigma_1 \sigma_2 \sigma_3\>-\<\sigma_1 \sigma_2\>\<\sigma_3\>-\<\sigma_1 \sigma_3\>\<\sigma_2\>-\<\sigma_2 \sigma_3\>\<\sigma_1\>+2 \<\sigma_1\>\<\sigma_2\>\<\sigma_3\>$, as a function of
   the pairwise correlation between 3 neurons firing with probability
   $\<\sigma_1\>=\<\sigma_2\>=\<\sigma_3\>=0.02$ (mean empirical value). 
Pairwise correlation in the retinal
data range from  $-10^{-3}$ to $0.03$ with a median of $2\cdot
10^{-4}$. Key is as in (a).
\label{fig2}
}
\end{center}
\end{figure}

\begin{figure}
\begin{center}
\noindent\includegraphics[width=\linewidth]{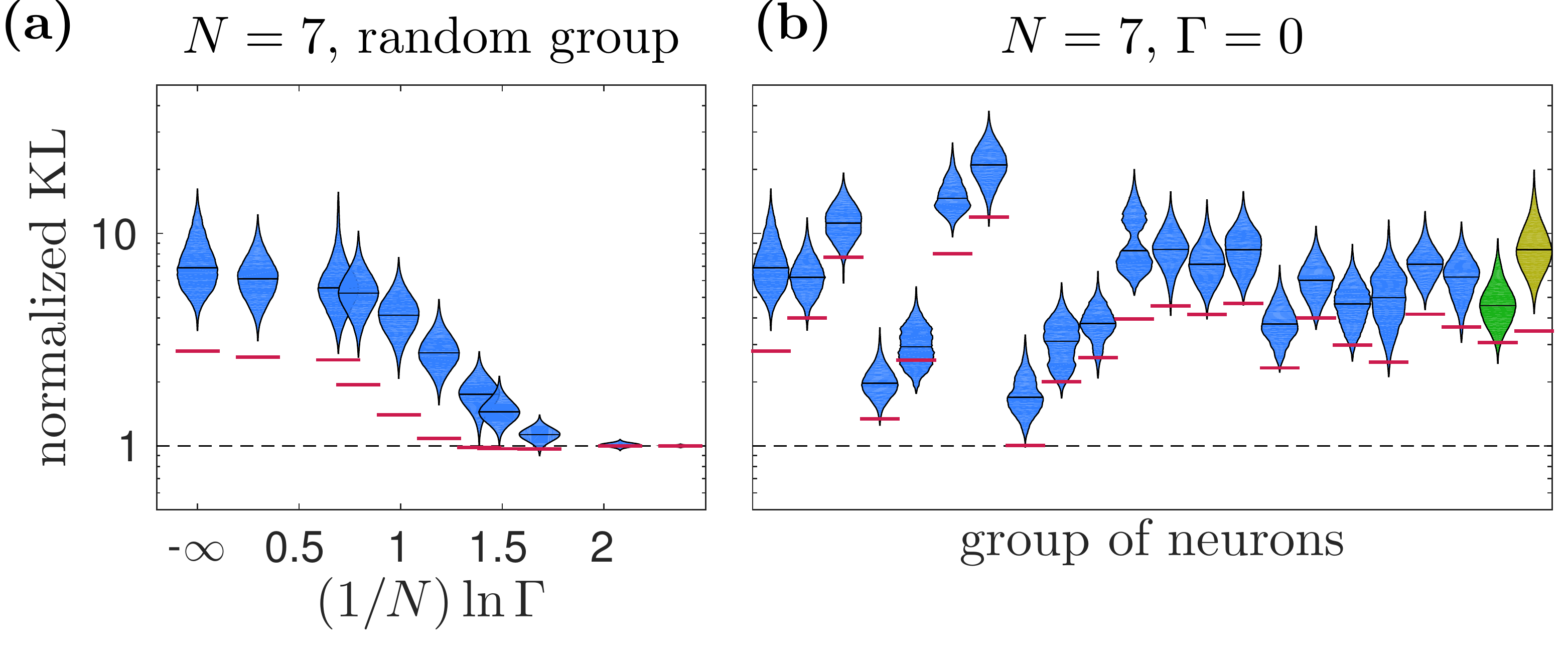}
\caption{{\bf Random versus maximum entropy models.} (a) The normalized Kullback-Leibler divergence relative to maximum entropy, $D_{\rm KL}(\hat P\Vert P)/D_{\rm KL}(\hat P\Vert P_{\rm ME})$,
 is represented as a function of the entropy-conjugated variable $\Gamma$ for (a) a random group of $N=7$ neurons (out of 40). 
Values above unity (dashed line) mean that maximum entropy outperforms the random model. 
The violin plots show the distributions over random models drawn from
$\mu_\Gamma$, while the red lines show the value for the average model, $D_{\rm KL}(\hat P\Vert P_\Gamma)$.
(b) Normalized KL divergence at $\Gamma=0$ for 20 random subsets of 7 neurons (blue), as well as the group of most correlated neurons (as measured by Pearson's correlation coefficient, green), and the set of neurons with the highest spike rate (yellow).
\label{fig3}
}
\end{center}
\end{figure}

We follow the approach of the random ensemble defined by
(\ref{mu}) to describe the joint spiking activity of retinal ganglion cells reported in \cite{Schneidman2006}. 
There, the spiking activities of 40 ganglion cells from the salamander retina
were recorded by multielectrode arrays for about an
hour, and segmented into $\approx 1.5\cdot 10^5$ binary spike words
$\sigma$ of $20$ ms.
The collective behavior
of small networks (up to 10 neurons) was shown to be well described by maximum
entropy distributions constrained by spike rates and pairwise
correlations (and later to much larger populations \cite{Tkacik2014a}).
This choice of constraints corresponds to the observables
$\mathcal{O}_a=\sigma_i$ for all neuron $i$, and
$\mathcal{O}_a=\sigma_i\sigma_j$ for all pairs $i,j$, for which the
maximum entropy distribution \eqref{maxent} takes the form of
a disordered Ising model, 
$P_{\rm ME}(\sigma)=(1/Z)\exp(\sum_i h_i\sigma_i+\sum_{ij}J_{ij}\sigma_i\sigma_j)$.

It is instructive first to consider the unbiased measure $\mu_{0}$
over very small networks, for which everything can be calculated analytically. The simplest case of two neurons constrained by
just their firing rate is illustrated by Fig.~\ref{fig2}a. The
maximum entropy distribution factorizes over the two neurons, which
are thus independent \cite{Schneidman2003}: $P(\sigma)=p_1(\sigma_1)p_2(\sigma_2)$. By constrast,
random models drawn from $\mu_0$ are biased towards a positive
correlation $\<\sigma_1\sigma_2\>-\<\sigma_1\>\<\sigma_2\>>0$ 
when both firing rates
$\<\sigma_1\>,\<\sigma_2\>$ are on the same side of $0.5$ (in the
retinal data $\<\sigma_i\>\sim 0.02$).
A similar bias in
the triplet correlation is
also found when considering 3 neurons constrained by uniform firing rates and
pairwise correlations (Fig.~\ref{fig2}b). When pairwise
correlations are weak, as is the case in the retina \cite{Schneidman2006},
random models predict on average a higher 3-point connected
correlation than maximum entropy, although the bias is reversed
for large correlations. 

Thanks to its exponential form \eqref{maxent}, the maximum entropy
distribution can be inferred with relative ease for systems of size $N\leq 20$,
yet requiring to calculate sums of $2^N$ terms \cite{Schneidman2006}.
Sampling from
$\mu_\Gamma$ or calculating $P_{\Gamma}$, on the other hand, is a much harder task, involving the exploration of $\mathcal{C}$ of dimension $2^N-N(N+1)/2-1$.
To apply the random ensemble to populations of neurons, we sampled
from $\mu_\Gamma$ using the Metropolis-Hastings algorithm, for various subgroups of neurons of different sizes.
At each step, starting from a distribution $P$ in $\mathcal{C}$, one
picks a random direction $V$ in the Fourier basis of the hyperplane
orthogonal to all observables $\mathcal{O}_a$  \cite{Obuchi2015}. The new distribution is
taken to be $P'=P+\alpha V$, where $\alpha$ is drawn uniformly in the
interval $(\alpha_{\rm min},\alpha_{\rm max})$ defined by the lower
and upper limits so that $P'(\sigma)\geq 0$ for all $\sigma$. $P'$ is accepted with probability $\min(1,e^{\Gamma [S(P')-S(P)]})$. The process is repeated until equilibration is reached.
High space dimension
limits us to relatively small group sizes, $N\leq 8$. Fortunately for
these sizes the true distribution $\hat P$ may be accurately estimated
from the data, and directly compared to models.

The accuracy of a given model is assessed by the
Kullback-Leibler (KL) divergence between the model distribution $P$ and the
true one $\hat P$, $D_{\rm KL}(\hat P\Vert P)=\sum_\sigma
\hat P(\sigma)\ln [\hat P(\sigma)/P(\sigma)]$. 
Fig.~\ref{fig3} shows, in the form of violin plots, the
distribution of KL divergence (normalized relative to maximum
entropy) when sampling $P$ from $\mu_\Gamma$, for
groups of $N=7$ cells. This distribution is plotted in Fig.~\ref{fig3}a for a random groups of 7 cells. Maximum entropy is found to have a clear advantage: 
its accuracy is matched by only a negligible fraction of models drawn
from $\mu_{\Gamma}$, and it also does better than their mean $P_\Gamma$
(red line).
The advantage of maximum entropy over the unbiased ensemble
generalizes to 20 random groups of 7 cells (Fig.~\ref{fig3}b), as well as the groups comprising the most correlated (green) and most active (yellow) cells.
Interestingly, in all cases the mean distribution $P_\Gamma$ is more accurate than the typical distribution $P$ sampled from $\mu_\Gamma$, a consequence of Jensen's inequality which implies
$D_{\rm  KL}(\hat P\Vert \<P\>_\Gamma)\leq \<D_{\rm  KL}(\hat P\Vert P)\>_\Gamma$.
In general, $0<\Gamma < \infty$ interpolates between the unbiased ensemble and the maximum entropy distribution.
For these reasons, in the following the maximum entropy model $P_{\rm ME}$ will only be compared to the mean distribution of the unbiased ensemble, $P_0$.

\begin{figure}
\begin{center}
\noindent\includegraphics[width=.49\linewidth]{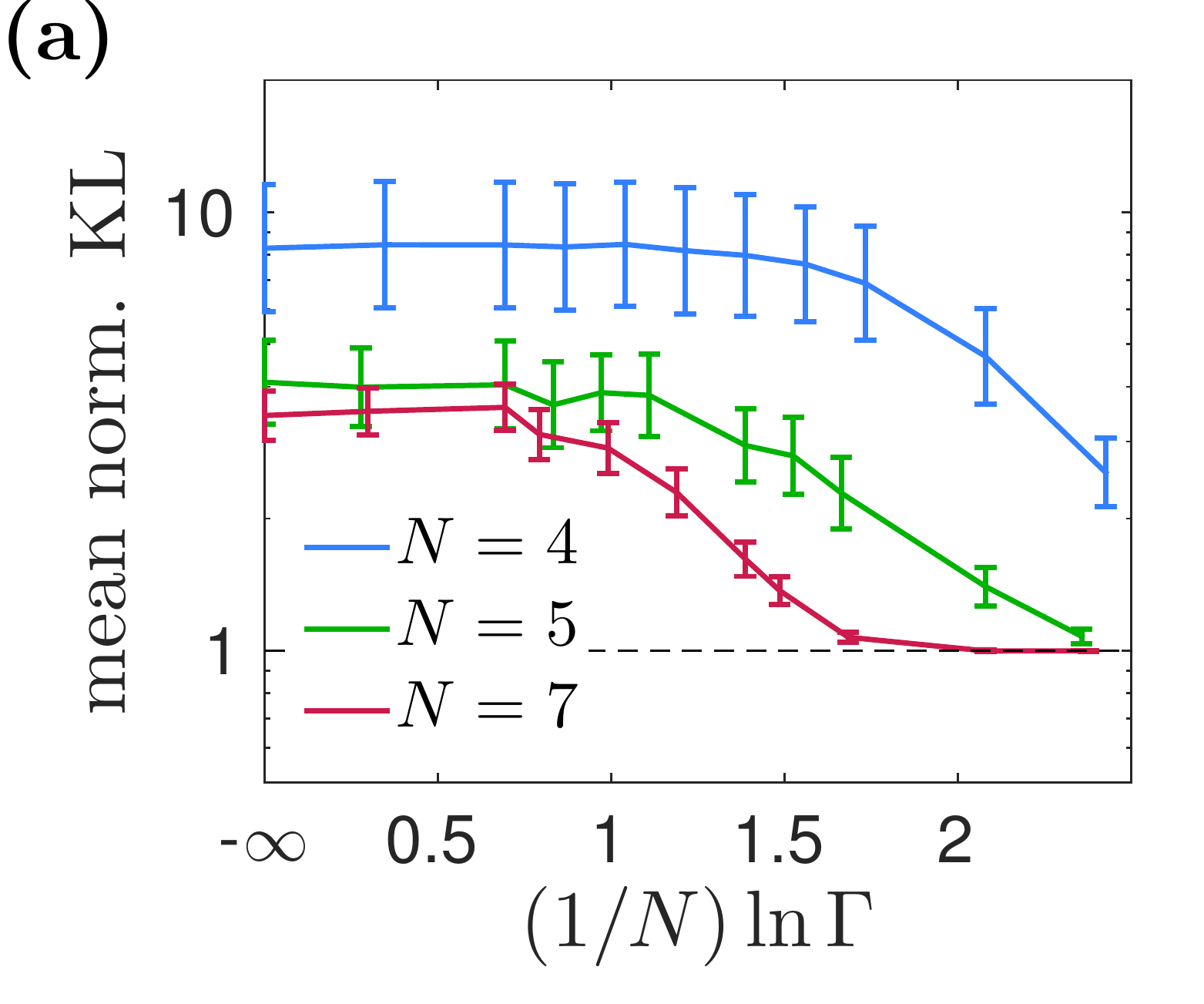}
\noindent\includegraphics[width=.49\linewidth]{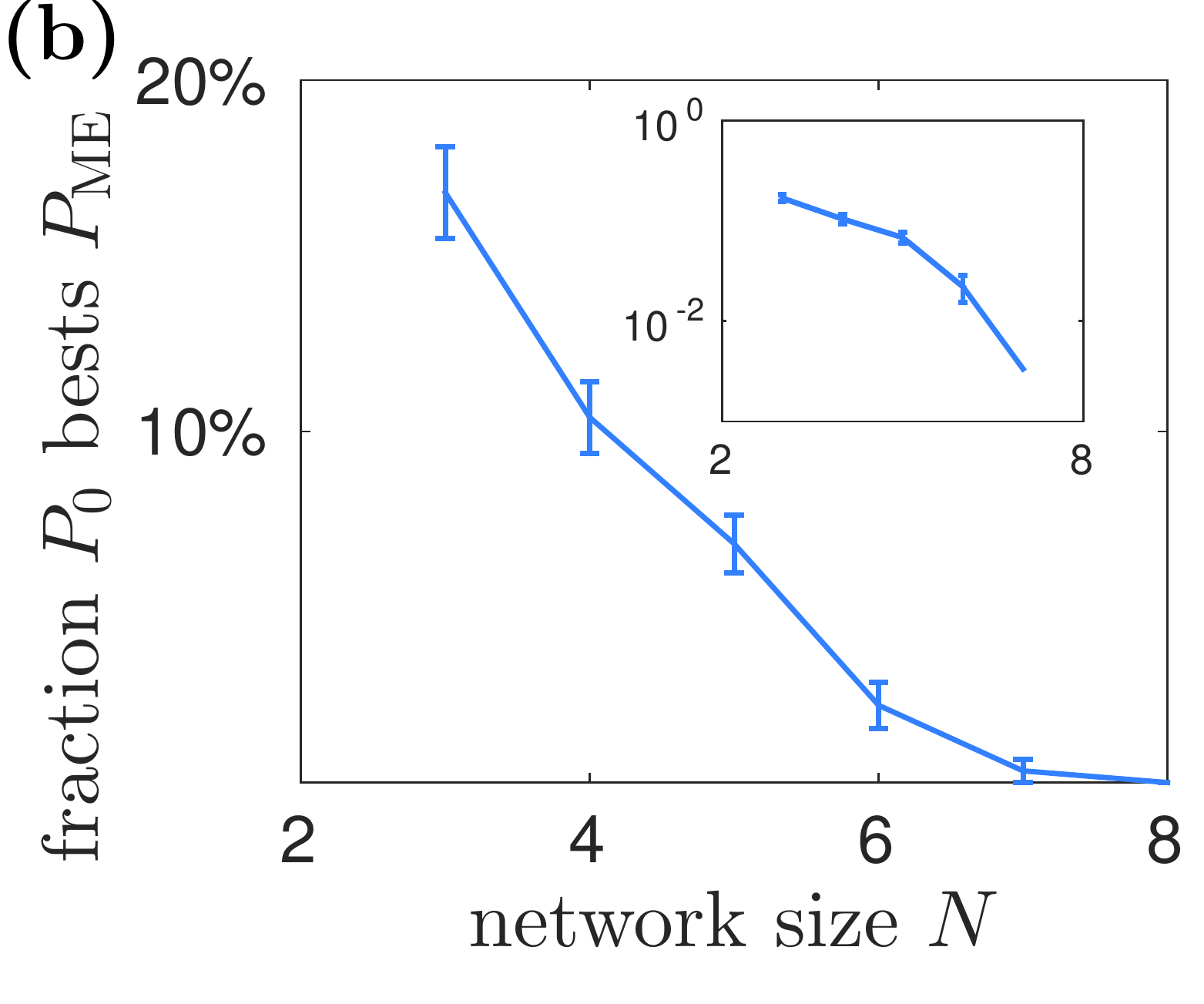}
\caption{{\bf Dependence on populations size.} (a) The normalized divergence
  of the average model, $D_{\rm KL}(\hat P\Vert P_\Gamma)$, is
  averaged over 20 random subsets, and plotted as a function of
  $(1/N)\ln \Gamma$. Errors bars show standard error on the mean. (b)
  Fraction of random groups (out of hundreds) of $N$ neurons that are better described
  by the mean unbiased distribution $P_0$ than by the maximum entropy
  model $P_{\rm ME}$.
\label{fig4}
}
\end{center}
\end{figure}

We now investigate the dependence on the population size.
Fig.~\ref{fig4} shows the average normalized KL divergence of the mean
model $P_\Gamma$ for random cell groups of varying sizes, as a
function of $(1/N)\ln\Gamma$ (the scaling of $\Gamma$ is assumed to be
exponential in $N$, as suggested by calculations with random
observables \cite{Obuchi2015a}). The general trend noted before for $N=7$ generalizes to all sizes: the larger the entropy bias $\Gamma$,
the better the model (Fig.~\ref{fig4}a). However, this average
behaviour masks large heterogeneities across different choices of cell
groups, especially for small groups, of which a sizeable fraction is
better described by the mean distribution $P_0$ than by $P_{\rm ME}$. Evaluating this fraction from hundreds of random groups for each $N$, we find that
maximum entropy is more likely to outperform the random ensemble in larger groups (Fig.~\ref{fig4}b), and even does so in all of the 200 tested groups of size $N=8$.

\begin{figure*}
\begin{center}
\noindent\includegraphics[width=.66\linewidth]{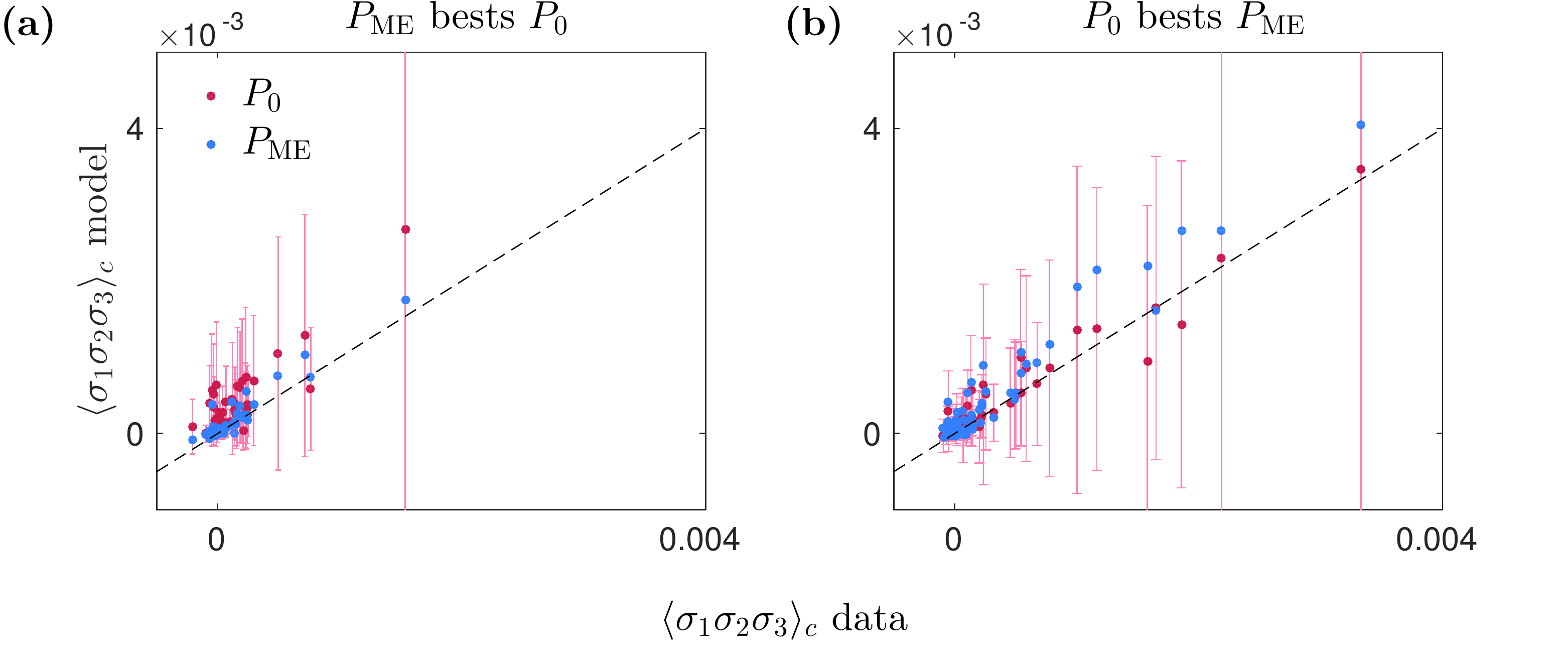}
\noindent\includegraphics[width=.33\linewidth]{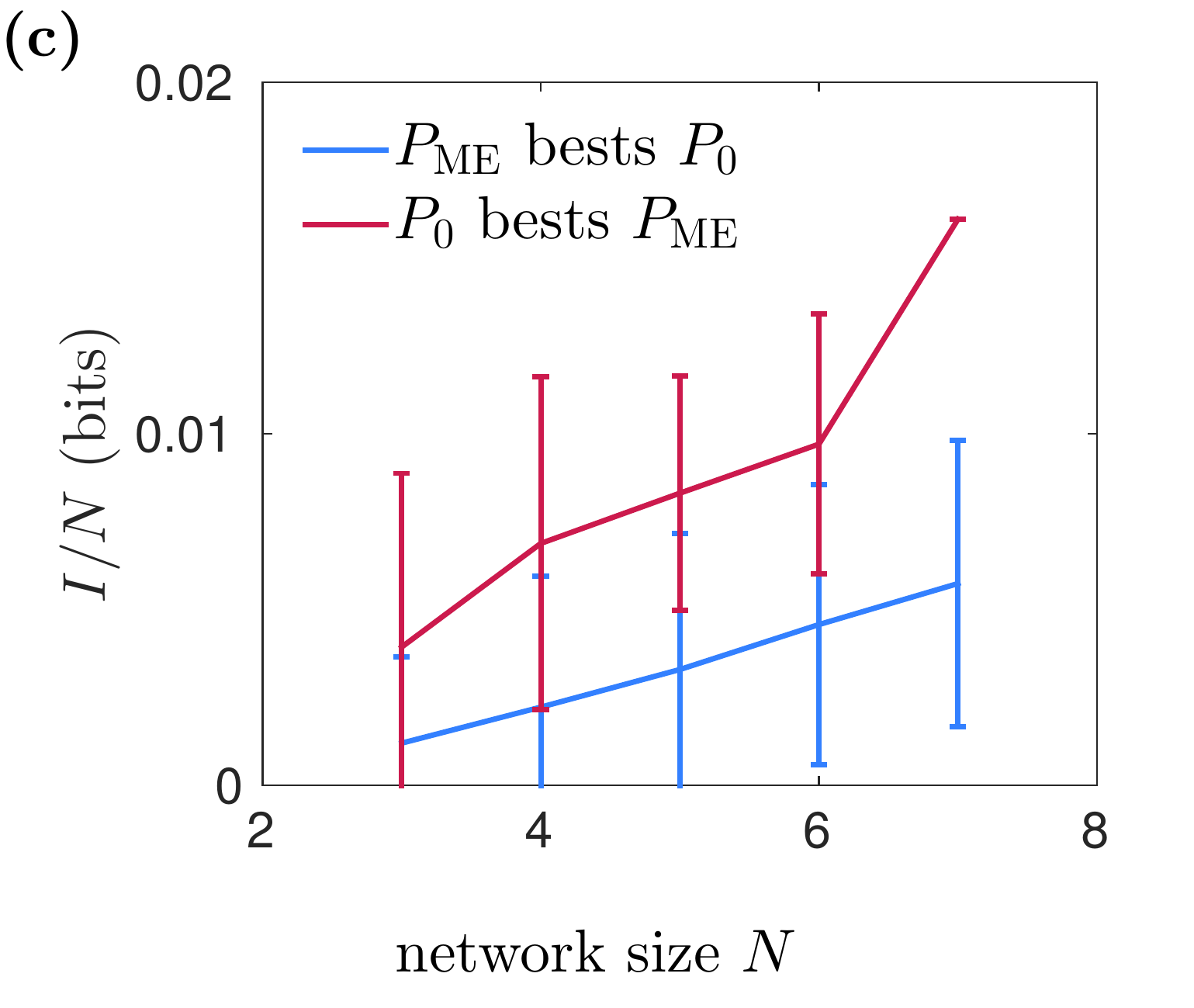}
\caption{{\bf Correlations and maximum likelihood performance.} Three-point connected correlation $\<\sigma_1\sigma_2\sigma_3\>_c$ for (a) 100 random triplets whose joint activity is best described by maximum entropy and (b) 100 random triplets whose joint activity is best described by the mean unbiased model, when constraining the values of the pairwise correlations. The error bar shows, for each triplet, the allowed range of values for the 3-point correlation.
(c) The multi-information, which measures the overall amount of correlation in the  collective acitivity, is plotted as a function of system size, for groups of
  neurons that are best described by the maximum entropy model $P_{\rm ME}$ (blue) or   by the mean unbiased model $P_0$ (red). Error bars show standard deviation  across groups of cells (the red point at $N=7$ has no error bar because only one group of that size was better described by $P_0$)
\label{fig5}
}
\end{center}
\end{figure*}

What sets apart groups of cells that are better described by $P_0$ than by $P_{\rm ME}$?
Since both share the same 1- and 2-point correlations by construction, we examine their predictions for 3-point correlations in triplets of cells ($N=3$). Fig.~\ref{fig5}a shows that random models typically fail because they overestimate small 3-point correlations. By contrast, maximum entropy is more likely to be outperformed by random models when the triplet correlation is large, in which case maximum entropy overestimates it. Both these findings are in agreement with the results of Fig.~\ref{fig2}b.
This observation can be generalized to larger groups of neurons ($N>3$)
by considering
the total amount of correlations in the network, quantified by
the loss of entropy due to correlations, or multi-information \cite{Schneidman2003},
$I=S(P_{\rm
  ind}) - S(\hat P)$,
where $P_{\rm ind}=\prod_i p_i(\sigma_i)$ is the model
distribution of independent neurons.
Groups that are better
described by $P_0$ than by $P_{\rm ME}$ are
found to have a higher multi-information on average (Fig.~\ref{fig5}c).

Since maximum entropy was proposed as a method for building statistical models from high-dimensional data, its accuracy, relevance, and epistemological validity have been questioned.
In this study we have shown that the maximum entropy model describes the
spiking activity of populations in the retina better
than the mean
model satisfying the same constraints, which itself performs better
than the vast majority of random models under these constraints.
This better performance of maximum entropy gets more marked as the population size $N$ grows, and is essentially always true for  $N\geq 8$.
The analysis of 3-point and higher-order correlations suggests that the rare instances where the mean model outperforms maximum entropy is when correlations are relatively large. In that case, maximum entropy predicts high triplet correlations within the allowed range compared to the mean unbiased model (Fig.~\ref{fig2}b), and may thus overestimate their true value, consistent with previous observations in large populations \cite{Tkacik2014a}. In that case, models that take a ``middle-of-the-road'' value of the correlations may be preferred to maximum entropy.

By providing a test on empirical data, our results complement previous
work aimed at explaining or refuting the efficiency of
maximum entropy based on theoretical arguments and simulated datasets.
Calculations on synthetic learning problems have suggested
that maximum entropy is no more accurate than random
\cite{Obuchi2015a}, unless the chosen observables are smooth as a
function of configuration space \cite{Obuchi2015}. However, in these
studies the choice of observables and true distributions were taken to
be completely random, and it is not clear how applicable they
are to real distributions and to pairwise constraints. Other simulation
studies have more specifically addressed the role of pairwise
interactions. It was suggested that pairwise maximum entropy models
should fail for large populations \cite{Roudi2009}. On the other hand, strongly
interacting systems with interactions of arbitrary order have been numerically
shown to be well described by pairwise interactions, with an analogy to
Hopfield networks \cite{Merchan2016}.
The principle of maximum entropy has also been advocated by
contrast to non-additive (or R\'enyi) entropies, but on purely theoretical
grounds \cite{Presse2013b}. Our results do not preclude that other
objective functions
than entropy may help better describe empirical data. They suggest, however,
that it is better to pick the most 
random model than to pick a model at random.

We thank Michael Berry for sharing the data from \cite{Schneidman2006}, and Olivier Marre for his comments on the manuscript.


\begin{thebibliography}{32}
\expandafter\ifx\csname natexlab\endcsname\relax\def\natexlab#1{#1}\fi
\expandafter\ifx\csname bibnamefont\endcsname\relax
  \def\bibnamefont#1{#1}\fi
\expandafter\ifx\csname bibfnamefont\endcsname\relax
  \def\bibfnamefont#1{#1}\fi
\expandafter\ifx\csname citenamefont\endcsname\relax
  \def\citenamefont#1{#1}\fi
\expandafter\ifx\csname url\endcsname\relax
  \def\url#1{\texttt{#1}}\fi
\expandafter\ifx\csname urlprefix\endcsname\relax\def\urlprefix{URL }\fi
\providecommand{\bibinfo}[2]{#2}
\providecommand{\eprint}[2][]{\url{#2}}

\bibitem[{\citenamefont{Jaynes}(1957{\natexlab{a}})}]{Jaynes1957a}
\bibinfo{author}{\bibfnamefont{E.~T.} \bibnamefont{Jaynes}},
  \bibinfo{journal}{Phys. Rev.} \textbf{\bibinfo{volume}{106}},
  \bibinfo{pages}{620} (\bibinfo{year}{1957}{\natexlab{a}}).

\bibitem[{\citenamefont{Jaynes}(1957{\natexlab{b}})}]{Jaynes1957}
\bibinfo{author}{\bibfnamefont{E.~T.} \bibnamefont{Jaynes}},
  \bibinfo{journal}{Phys. Rev.} \textbf{\bibinfo{volume}{108}},
  \bibinfo{pages}{171} (\bibinfo{year}{1957}{\natexlab{b}}).

\bibitem[{\citenamefont{Schneidman et~al.}(2006)\citenamefont{Schneidman,
  Berry, Segev, and Bialek}}]{Schneidman2006}
\bibinfo{author}{\bibfnamefont{E.}~\bibnamefont{Schneidman}},
  \bibinfo{author}{\bibfnamefont{M.~J.} \bibnamefont{Berry}},
  \bibinfo{author}{\bibfnamefont{R.}~\bibnamefont{Segev}}, \bibnamefont{and}
  \bibinfo{author}{\bibfnamefont{W.}~\bibnamefont{Bialek}},
  \bibinfo{journal}{Nature} \textbf{\bibinfo{volume}{440}},
  \bibinfo{pages}{1007} (\bibinfo{year}{2006}).

\bibitem[{\citenamefont{Shlens et~al.}(2006)\citenamefont{Shlens, Field,
  Gauthier, Grivich, Petrusca, Sher, Litke, and Chichilnisky}}]{Shlens2006}
\bibinfo{author}{\bibfnamefont{J.}~\bibnamefont{Shlens}},
  \bibinfo{author}{\bibfnamefont{G.~D.} \bibnamefont{Field}},
  \bibinfo{author}{\bibfnamefont{J.~L.} \bibnamefont{Gauthier}},
  \bibinfo{author}{\bibfnamefont{M.~I.} \bibnamefont{Grivich}},
  \bibinfo{author}{\bibfnamefont{D.}~\bibnamefont{Petrusca}},
  \bibinfo{author}{\bibfnamefont{A.}~\bibnamefont{Sher}},
  \bibinfo{author}{\bibfnamefont{A.~M.} \bibnamefont{Litke}}, \bibnamefont{and}
  \bibinfo{author}{\bibfnamefont{E.~J.} \bibnamefont{Chichilnisky}},
  \bibinfo{journal}{J. Neurosci.} \textbf{\bibinfo{volume}{26}},
  \bibinfo{pages}{8254} (\bibinfo{year}{2006}).

\bibitem[{\citenamefont{Tang et~al.}(2008)\citenamefont{Tang, Jackson, Hobbs,
  Chen, Smith, Patel, Prieto, Petrusca, Grivich, Sher et~al.}}]{Tang2008}
\bibinfo{author}{\bibfnamefont{A.}~\bibnamefont{Tang}},
  \bibinfo{author}{\bibfnamefont{D.}~\bibnamefont{Jackson}},
  \bibinfo{author}{\bibfnamefont{J.}~\bibnamefont{Hobbs}},
  \bibinfo{author}{\bibfnamefont{W.}~\bibnamefont{Chen}},
  \bibinfo{author}{\bibfnamefont{J.~L.} \bibnamefont{Smith}},
  \bibinfo{author}{\bibfnamefont{H.}~\bibnamefont{Patel}},
  \bibinfo{author}{\bibfnamefont{A.}~\bibnamefont{Prieto}},
  \bibinfo{author}{\bibfnamefont{D.}~\bibnamefont{Petrusca}},
  \bibinfo{author}{\bibfnamefont{M.~I.} \bibnamefont{Grivich}},
  \bibinfo{author}{\bibfnamefont{A.}~\bibnamefont{Sher}}, \bibnamefont{et~al.},
  \bibinfo{journal}{J. Neurosci.} \textbf{\bibinfo{volume}{28}},
  \bibinfo{pages}{505} (\bibinfo{year}{2008}).

\bibitem[{\citenamefont{Tavoni et~al.}(2015)\citenamefont{Tavoni, Ferrari,
  Battaglia, and Cocco}}]{Tavoni2015}
\bibinfo{author}{\bibfnamefont{G.}~\bibnamefont{Tavoni}},
  \bibinfo{author}{\bibfnamefont{U.}~\bibnamefont{Ferrari}},
  \bibinfo{author}{\bibfnamefont{F.~P.} \bibnamefont{Battaglia}},
  \bibnamefont{and} \bibinfo{author}{\bibfnamefont{S.}~\bibnamefont{Cocco}}
  (\bibinfo{year}{2015}).

\bibitem[{\citenamefont{Watanabe et~al.}(2013)\citenamefont{Watanabe, Hirose,
  Wada, Imai, Machida, Shirouzu, Konishi, Miyashita, and
  Masuda}}]{Watanabe2013a}
\bibinfo{author}{\bibfnamefont{T.}~\bibnamefont{Watanabe}},
  \bibinfo{author}{\bibfnamefont{S.}~\bibnamefont{Hirose}},
  \bibinfo{author}{\bibfnamefont{H.}~\bibnamefont{Wada}},
  \bibinfo{author}{\bibfnamefont{Y.}~\bibnamefont{Imai}},
  \bibinfo{author}{\bibfnamefont{T.}~\bibnamefont{Machida}},
  \bibinfo{author}{\bibfnamefont{I.}~\bibnamefont{Shirouzu}},
  \bibinfo{author}{\bibfnamefont{S.}~\bibnamefont{Konishi}},
  \bibinfo{author}{\bibfnamefont{Y.}~\bibnamefont{Miyashita}},
  \bibnamefont{and} \bibinfo{author}{\bibfnamefont{N.}~\bibnamefont{Masuda}},
  \bibinfo{journal}{Nat. Commun.} \textbf{\bibinfo{volume}{4}},
  \bibinfo{pages}{1370} (\bibinfo{year}{2013}).

\bibitem[{\citenamefont{Weigt et~al.}(2009)\citenamefont{Weigt, White,
  Szurmant, Hoch, and Hwa}}]{Weigt2009}
\bibinfo{author}{\bibfnamefont{M.}~\bibnamefont{Weigt}},
  \bibinfo{author}{\bibfnamefont{R.~a.} \bibnamefont{White}},
  \bibinfo{author}{\bibfnamefont{H.}~\bibnamefont{Szurmant}},
  \bibinfo{author}{\bibfnamefont{J.~a.} \bibnamefont{Hoch}}, \bibnamefont{and}
  \bibinfo{author}{\bibfnamefont{T.}~\bibnamefont{Hwa}},
  \bibinfo{journal}{Proc. Natl. Acad. Sci. U. S. A.}
  \textbf{\bibinfo{volume}{106}}, \bibinfo{pages}{67} (\bibinfo{year}{2009}).

\bibitem[{\citenamefont{Mora et~al.}(2010)\citenamefont{Mora, Walczak, Bialek,
  and Callan}}]{Mora2010}
\bibinfo{author}{\bibfnamefont{T.}~\bibnamefont{Mora}},
  \bibinfo{author}{\bibfnamefont{A.~M.} \bibnamefont{Walczak}},
  \bibinfo{author}{\bibfnamefont{W.}~\bibnamefont{Bialek}}, \bibnamefont{and}
  \bibinfo{author}{\bibfnamefont{C.~G.} \bibnamefont{Callan}},
  \bibinfo{journal}{Proc. Natl. Acad. Sci.} \textbf{\bibinfo{volume}{107}},
  \bibinfo{pages}{5405} (\bibinfo{year}{2010}), \eprint{0912.5175}.

\bibitem[{\citenamefont{Figliuzzi et~al.}(2016)\citenamefont{Figliuzzi,
  Jacquier, Schug, Tenaillon, and Weigt}}]{Figliuzzi2016}
\bibinfo{author}{\bibfnamefont{M.}~\bibnamefont{Figliuzzi}},
  \bibinfo{author}{\bibfnamefont{H.}~\bibnamefont{Jacquier}},
  \bibinfo{author}{\bibfnamefont{A.}~\bibnamefont{Schug}},
  \bibinfo{author}{\bibfnamefont{O.}~\bibnamefont{Tenaillon}},
  \bibnamefont{and} \bibinfo{author}{\bibfnamefont{M.}~\bibnamefont{Weigt}},
  \bibinfo{journal}{Mol. Biol. Evol.} \textbf{\bibinfo{volume}{33}},
  \bibinfo{pages}{268} (\bibinfo{year}{2016}), \eprint{1510.03224}.

\bibitem[{\citenamefont{Santolini et~al.}(2014)\citenamefont{Santolini, Mora,
  and Hakim}}]{Santolini2014}
\bibinfo{author}{\bibfnamefont{M.}~\bibnamefont{Santolini}},
  \bibinfo{author}{\bibfnamefont{T.}~\bibnamefont{Mora}}, \bibnamefont{and}
  \bibinfo{author}{\bibfnamefont{V.}~\bibnamefont{Hakim}},
  \bibinfo{journal}{PLoS One} \textbf{\bibinfo{volume}{9}},
  \bibinfo{pages}{e99015} (\bibinfo{year}{2014}).

\bibitem[{\citenamefont{{De Leonardis} et~al.}(2015)\citenamefont{{De
  Leonardis}, Lutz, Ratz, Cocco, Monasson, Schug, and Weigt}}]{DeLeonardis2015}
\bibinfo{author}{\bibfnamefont{E.}~\bibnamefont{{De Leonardis}}},
  \bibinfo{author}{\bibfnamefont{B.}~\bibnamefont{Lutz}},
  \bibinfo{author}{\bibfnamefont{S.}~\bibnamefont{Ratz}},
  \bibinfo{author}{\bibfnamefont{S.}~\bibnamefont{Cocco}},
  \bibinfo{author}{\bibfnamefont{R.}~\bibnamefont{Monasson}},
  \bibinfo{author}{\bibfnamefont{A.}~\bibnamefont{Schug}}, \bibnamefont{and}
  \bibinfo{author}{\bibfnamefont{M.}~\bibnamefont{Weigt}},
  \bibinfo{journal}{Nucleic Acids Res.} \textbf{\bibinfo{volume}{43}},
  \bibinfo{pages}{10444} (\bibinfo{year}{2015}).

\bibitem[{\citenamefont{Bialek et~al.}(2012)\citenamefont{Bialek, Cavagna,
  Giardina, Mora, Silvestri, Viale, and Walczak}}]{Bialek2012}
\bibinfo{author}{\bibfnamefont{W.}~\bibnamefont{Bialek}},
  \bibinfo{author}{\bibfnamefont{A.}~\bibnamefont{Cavagna}},
  \bibinfo{author}{\bibfnamefont{I.}~\bibnamefont{Giardina}},
  \bibinfo{author}{\bibfnamefont{T.}~\bibnamefont{Mora}},
  \bibinfo{author}{\bibfnamefont{E.}~\bibnamefont{Silvestri}},
  \bibinfo{author}{\bibfnamefont{M.}~\bibnamefont{Viale}}, \bibnamefont{and}
  \bibinfo{author}{\bibfnamefont{A.~M.} \bibnamefont{Walczak}},
  \bibinfo{journal}{Proc. Natl. Acad. Sci. U. S. A.}
  \textbf{\bibinfo{volume}{109}}, \bibinfo{pages}{4786} (\bibinfo{year}{2012}),
  \eprint{1107.0604}.

\bibitem[{\citenamefont{Stephens and Bialek}(2010)}]{Stephens2010}
\bibinfo{author}{\bibfnamefont{G.~J.} \bibnamefont{Stephens}} \bibnamefont{and}
  \bibinfo{author}{\bibfnamefont{W.}~\bibnamefont{Bialek}},
  \bibinfo{journal}{Phys. Rev. E - Stat. Nonlinear, Soft Matter Phys.}
  \textbf{\bibinfo{volume}{81}}, \bibinfo{pages}{3} (\bibinfo{year}{2010}),
  \eprint{0801.0253}.

\bibitem[{\citenamefont{Lee et~al.}(2015)\citenamefont{Lee, Broedersz, and
  Bialek}}]{Lee2015}
\bibinfo{author}{\bibfnamefont{E.~D.} \bibnamefont{Lee}},
  \bibinfo{author}{\bibfnamefont{C.~P.} \bibnamefont{Broedersz}},
  \bibnamefont{and} \bibinfo{author}{\bibfnamefont{W.}~\bibnamefont{Bialek}},
  \bibinfo{journal}{J. Stat. Phys.} \textbf{\bibinfo{volume}{160}},
  \bibinfo{pages}{275} (\bibinfo{year}{2015}), \eprint{1306.5004}.

\bibitem[{\citenamefont{Mora and Bialek}(2011)}]{Mora2011a}
\bibinfo{author}{\bibfnamefont{T.}~\bibnamefont{Mora}} \bibnamefont{and}
  \bibinfo{author}{\bibfnamefont{W.}~\bibnamefont{Bialek}},
  \bibinfo{journal}{J. Stat. Phys.} \textbf{\bibinfo{volume}{144}},
  \bibinfo{pages}{268} (\bibinfo{year}{2011}).

\bibitem[{\citenamefont{Bialek et~al.}(2014)\citenamefont{Bialek, Cavagna,
  Giardina, Mora, Pohl, Silvestri, Viale, and Walczak}}]{Bialek2014}
\bibinfo{author}{\bibfnamefont{W.}~\bibnamefont{Bialek}},
  \bibinfo{author}{\bibfnamefont{A.}~\bibnamefont{Cavagna}},
  \bibinfo{author}{\bibfnamefont{I.}~\bibnamefont{Giardina}},
  \bibinfo{author}{\bibfnamefont{T.}~\bibnamefont{Mora}},
  \bibinfo{author}{\bibfnamefont{O.}~\bibnamefont{Pohl}},
  \bibinfo{author}{\bibfnamefont{E.}~\bibnamefont{Silvestri}},
  \bibinfo{author}{\bibfnamefont{M.}~\bibnamefont{Viale}}, \bibnamefont{and}
  \bibinfo{author}{\bibfnamefont{A.~M.} \bibnamefont{Walczak}},
  \bibinfo{journal}{Proc. Natl. Acad. Sci. U. S. A.}
  \textbf{\bibinfo{volume}{111}}, \bibinfo{pages}{7212} (\bibinfo{year}{2014}).

\bibitem[{\citenamefont{Tka{\v{c}}ik et~al.}(2015)\citenamefont{Tka{\v{c}}ik,
  Mora, Marre, Amodei, Palmer, Berry, and Bialek}}]{Tkacik2015}
\bibinfo{author}{\bibfnamefont{G.}~\bibnamefont{Tka{\v{c}}ik}},
  \bibinfo{author}{\bibfnamefont{T.}~\bibnamefont{Mora}},
  \bibinfo{author}{\bibfnamefont{O.}~\bibnamefont{Marre}},
  \bibinfo{author}{\bibfnamefont{D.}~\bibnamefont{Amodei}},
  \bibinfo{author}{\bibfnamefont{S.~E.} \bibnamefont{Palmer}},
  \bibinfo{author}{\bibfnamefont{M.~J.} \bibnamefont{Berry}}, \bibnamefont{and}
  \bibinfo{author}{\bibfnamefont{W.}~\bibnamefont{Bialek}},
  \bibinfo{journal}{Proc. Natl. Acad. Sci.} \textbf{\bibinfo{volume}{112}},
  \bibinfo{pages}{11508} (\bibinfo{year}{2015}).

\bibitem[{\citenamefont{Watanabe et~al.}(1)\citenamefont{Watanabe, Masuda,
  Megumi, Kanai, and Rees}}]{Watanabe2014}
\bibinfo{author}{\bibfnamefont{T.}~\bibnamefont{Watanabe}},
  \bibinfo{author}{\bibfnamefont{N.}~\bibnamefont{Masuda}},
  \bibinfo{author}{\bibfnamefont{F.}~\bibnamefont{Megumi}},
  \bibinfo{author}{\bibfnamefont{R.}~\bibnamefont{Kanai}}, \bibnamefont{and}
  \bibinfo{author}{\bibfnamefont{G.}~\bibnamefont{Rees}},
  \bibinfo{journal}{Nat. Commun.} \textbf{\bibinfo{volume}{5}},
  \bibinfo{pages}{1} (\bibinfo{year}{1}), \eprint{arXiv:1011.1669v3}.

\bibitem[{\citenamefont{Tka{\v{c}}ik et~al.}(2014)\citenamefont{Tka{\v{c}}ik,
  Marre, Amodei, Schneidman, Bialek, and Berry}}]{Tkacik2014a}
\bibinfo{author}{\bibfnamefont{G.}~\bibnamefont{Tka{\v{c}}ik}},
  \bibinfo{author}{\bibfnamefont{O.}~\bibnamefont{Marre}},
  \bibinfo{author}{\bibfnamefont{D.}~\bibnamefont{Amodei}},
  \bibinfo{author}{\bibfnamefont{E.}~\bibnamefont{Schneidman}},
  \bibinfo{author}{\bibfnamefont{W.}~\bibnamefont{Bialek}}, \bibnamefont{and}
  \bibinfo{author}{\bibfnamefont{M.~J.} \bibnamefont{Berry}},
  \bibinfo{journal}{PLoS Comput. Biol.} \textbf{\bibinfo{volume}{10}},
  \bibinfo{pages}{e1003408} (\bibinfo{year}{2014}),
  \eprint{/arxiv.org/abs/1306.3061}.

\bibitem[{\citenamefont{Morcos et~al.}(2011)\citenamefont{Morcos, Pagnani,
  Lunt, Bertolino, Marks, Sander, Zecchina, Onuchic, Hwa, and
  Weigt}}]{Morcos2011}
\bibinfo{author}{\bibfnamefont{F.}~\bibnamefont{Morcos}},
  \bibinfo{author}{\bibfnamefont{A.}~\bibnamefont{Pagnani}},
  \bibinfo{author}{\bibfnamefont{B.}~\bibnamefont{Lunt}},
  \bibinfo{author}{\bibfnamefont{A.}~\bibnamefont{Bertolino}},
  \bibinfo{author}{\bibfnamefont{D.~S.} \bibnamefont{Marks}},
  \bibinfo{author}{\bibfnamefont{C.}~\bibnamefont{Sander}},
  \bibinfo{author}{\bibfnamefont{R.}~\bibnamefont{Zecchina}},
  \bibinfo{author}{\bibfnamefont{J.~N.} \bibnamefont{Onuchic}},
  \bibinfo{author}{\bibfnamefont{T.}~\bibnamefont{Hwa}}, \bibnamefont{and}
  \bibinfo{author}{\bibfnamefont{M.}~\bibnamefont{Weigt}},
  \bibinfo{journal}{Proc. Natl. Acad. Sci.} \textbf{\bibinfo{volume}{108}},
  \bibinfo{pages}{E1293} (\bibinfo{year}{2011}), \eprint{1110.5223}.

\bibitem[{\citenamefont{Ferguson et~al.}(2013)\citenamefont{Ferguson, Mann,
  Omarjee, Ndung'u, Walker, and Chakraborty}}]{Ferguson2013}
\bibinfo{author}{\bibfnamefont{A.~L.} \bibnamefont{Ferguson}},
  \bibinfo{author}{\bibfnamefont{J.~K.} \bibnamefont{Mann}},
  \bibinfo{author}{\bibfnamefont{S.}~\bibnamefont{Omarjee}},
  \bibinfo{author}{\bibfnamefont{T.}~\bibnamefont{Ndung'u}},
  \bibinfo{author}{\bibfnamefont{B.~D.} \bibnamefont{Walker}},
  \bibnamefont{and} \bibinfo{author}{\bibfnamefont{A.~K.}
  \bibnamefont{Chakraborty}}, \bibinfo{journal}{Immunity}
  \textbf{\bibinfo{volume}{38}}, \bibinfo{pages}{606} (\bibinfo{year}{2013}).

\bibitem[{\citenamefont{Shore and Johnson}(1980)}]{Shore1980}
\bibinfo{author}{\bibfnamefont{J.~E.} \bibnamefont{Shore}} \bibnamefont{and}
  \bibinfo{author}{\bibfnamefont{R.~W.} \bibnamefont{Johnson}},
  \bibinfo{journal}{IEEE Trans. Inf. Theory} \textbf{\bibinfo{volume}{26}},
  \bibinfo{pages}{26} (\bibinfo{year}{1980}).

\bibitem[{\citenamefont{Aurell}(2016)}]{Aurell2016}
\bibinfo{author}{\bibfnamefont{E.}~\bibnamefont{Aurell}},
  \bibinfo{journal}{PLoS Comput. Biol.} \textbf{\bibinfo{volume}{12}},
  \bibinfo{pages}{1} (\bibinfo{year}{2016}).

\bibitem[{\citenamefont{van Nimwegen}(2016)}]{VanNimwegen2016}
\bibinfo{author}{\bibfnamefont{E.}~\bibnamefont{van Nimwegen}},
  \bibinfo{journal}{PLoS Comput. Biol.} \textbf{\bibinfo{volume}{12}},
  \bibinfo{pages}{1} (\bibinfo{year}{2016}).

\bibitem[{\citenamefont{Obuchi et~al.}(2015)\citenamefont{Obuchi, Cocco, and
  Monasson}}]{Obuchi2015a}
\bibinfo{author}{\bibfnamefont{T.}~\bibnamefont{Obuchi}},
  \bibinfo{author}{\bibfnamefont{S.}~\bibnamefont{Cocco}}, \bibnamefont{and}
  \bibinfo{author}{\bibfnamefont{R.}~\bibnamefont{Monasson}},
  \bibinfo{journal}{J. Stat. Phys.} \textbf{\bibinfo{volume}{161}},
  \bibinfo{pages}{598} (\bibinfo{year}{2015}), \eprint{arXiv:1503.02802v1}.

\bibitem[{\citenamefont{Obuchi and Monasson}(2015)}]{Obuchi2015}
\bibinfo{author}{\bibfnamefont{T.}~\bibnamefont{Obuchi}} \bibnamefont{and}
  \bibinfo{author}{\bibfnamefont{R.}~\bibnamefont{Monasson}},
  \bibinfo{journal}{J. Phys. Conf. Ser.} \textbf{\bibinfo{volume}{638}},
  \bibinfo{pages}{012018} (\bibinfo{year}{2015}).

\bibitem[{\citenamefont{Presse et~al.}(2013)\citenamefont{Presse, Ghosh, Lee,
  and Dill}}]{Presse2013a}
\bibinfo{author}{\bibfnamefont{S.}~\bibnamefont{Presse}},
  \bibinfo{author}{\bibfnamefont{K.}~\bibnamefont{Ghosh}},
  \bibinfo{author}{\bibfnamefont{J.}~\bibnamefont{Lee}}, \bibnamefont{and}
  \bibinfo{author}{\bibfnamefont{K.~A.} \bibnamefont{Dill}},
  \bibinfo{journal}{Rev. Mod. Phys.} \textbf{\bibinfo{volume}{85}},
  \bibinfo{pages}{1115} (\bibinfo{year}{2013}).

\bibitem[{\citenamefont{Schneidman et~al.}(2003)\citenamefont{Schneidman,
  Still, Berry, and Bialek}}]{Schneidman2003}
\bibinfo{author}{\bibfnamefont{E.}~\bibnamefont{Schneidman}},
  \bibinfo{author}{\bibfnamefont{S.}~\bibnamefont{Still}},
  \bibinfo{author}{\bibfnamefont{M.~J.} \bibnamefont{Berry}}, \bibnamefont{and}
  \bibinfo{author}{\bibfnamefont{W.}~\bibnamefont{Bialek}},
  \bibinfo{journal}{Phys. Rev. Lett.} \textbf{\bibinfo{volume}{91}},
  \bibinfo{pages}{238701} (\bibinfo{year}{2003}).

\bibitem[{\citenamefont{Roudi et~al.}(2009)\citenamefont{Roudi, Nirenberg, and
  Latham}}]{Roudi2009}
\bibinfo{author}{\bibfnamefont{Y.}~\bibnamefont{Roudi}},
  \bibinfo{author}{\bibfnamefont{S.}~\bibnamefont{Nirenberg}},
  \bibnamefont{and} \bibinfo{author}{\bibfnamefont{P.~E.}
  \bibnamefont{Latham}}, \bibinfo{journal}{PLoS Comput. Biol.}
  \textbf{\bibinfo{volume}{5}}, \bibinfo{pages}{e1000380}
  (\bibinfo{year}{2009}).

\bibitem[{\citenamefont{Merchan and Nemenman}(2016)}]{Merchan2016}
\bibinfo{author}{\bibfnamefont{L.}~\bibnamefont{Merchan}} \bibnamefont{and}
  \bibinfo{author}{\bibfnamefont{I.}~\bibnamefont{Nemenman}},
  \bibinfo{journal}{J. Stat. Phys.} \textbf{\bibinfo{volume}{162}},
  \bibinfo{pages}{1294} (\bibinfo{year}{2016}), \eprint{1505.02831}.

\bibitem[{\citenamefont{Press{\'{e}} et~al.}(2013)\citenamefont{Press{\'{e}},
  Ghosh, Lee, and Dill}}]{Presse2013b}
\bibinfo{author}{\bibfnamefont{S.}~\bibnamefont{Press{\'{e}}}},
  \bibinfo{author}{\bibfnamefont{K.}~\bibnamefont{Ghosh}},
  \bibinfo{author}{\bibfnamefont{J.}~\bibnamefont{Lee}}, \bibnamefont{and}
  \bibinfo{author}{\bibfnamefont{K.~A.} \bibnamefont{Dill}},
  \bibinfo{journal}{Phys. Rev. Lett.} \textbf{\bibinfo{volume}{111}},
  \bibinfo{pages}{180604} (\bibinfo{year}{2013}), \eprint{arXiv:1312.1186v1}.

\end{thebibliography}
\end{document}